\documentstyle[preprint,revtex]{aps}
\def\bi{{\bf i}}

\begin{document}
\draft

%\widetext

\begin{title}
  Composite operators for BCS  Superconductor
\end{title}

\author{J.\ Bon\v ca}
\begin{instit}
Theoretical Division, T-11, Center for Nonlinear Studies, \\ Los
Alamos National Laboratory, Los Alamos, NM 87545 and \\
J. Stefan Institute, University of Ljubljana, 61111 Ljubljana, Slovenia
\end{instit}

\author{A. V. Balatsky}
\begin{instit}
Center for Materials Science and Theoretical Division, \\ Los
Alamos National Laboratory, Los Alamos, NM 87545 and\\ Landau
Institute for Theoretical Physics, Moscow, Russia
\end{instit}

\receipt{\today}

\vfill

\begin{abstract}
  The new form of the composite operator generalizing the Cooper pairs
for a BCS superconductor is introduced. The approach is similar to the
derivation of the composite operator of the odd - frequency
superconductors.  The examples of the $d_{x^2-y^2}-,~d_{xy}-$ and $p-$
wave composite operators for a 2D $t-J$ model are given.
\end{abstract}

Submitted to ZhETP Letters

\vfill


  Recently the notion of a composite operator as a generalization of
the Cooper pair in the theory of superconductivity has been proposed
\cite{a,bb,ek,c}. The examples of the composite operators are
\begin{equation} \Delta = \left\{ \begin{array}{ll}
    \langle \vec S ({\bf i})c_\alpha({\bf j})c_\beta({\bf k}) \rangle
    (i\sigma ^y\vec \sigma)_{\alpha\beta};~~~~ & \mbox{S=0}\\
    \langle \vec S ({\bf i})c_\alpha({\bf j})c_\beta({\bf k}) \rangle
    (\sigma ^x\vec \sigma)_{\alpha\beta};~~~~  & \mbox{$S=1,~S_z=0$}
\label{eq:1}
\end{array}
\right. ,
\end{equation}
which describe the bound state of the spin excitation at site $\bf i$
and a triplet or a singlet Cooper pair at $\bf j,~k$. Clearly these
operators carry charge 2e and thus describe the superconducting
ordering.  This type of the condensate is inherent for the
odd-frequency superconductors \cite{a,bb,ek,c}, which might occur in
strongly correlated systems such as t-J and Hubbard models. Indeed, it
has been argued that frustration of the magnetic degrees of freedom by
carriers may produce enhanced composite pairing correlations, for the
operators similar to Eq.  (1) \cite{bb}.

  The composite operator in the theory of superconductivity represents
a new level in the hierarchy of the possible superconducting
condensate. Any number of particle-hole ({\it i.e.} neutral) operators
composed together with the Cooper pair operator possess charge 2e and
thus can, in principle, describe some superconducting state
\cite{com}. Because of this general argument we point out that the
composite operators are possible for BCS superconductors as well as
for odd frequency ones.

The purpose of this note is to show that composite operators similar
to Eq. (\ref{eq:1}) can be constructed also in the case of BCS
pairing. As an example we will consider 2D t-J model on the square
lattice. Previously, the composite operator for particular case of
$d_{x^2-y^2}$ symmetry in a 2D t-J model was considered by Poilblanc
\cite{p}. The most relevant for possible BCS (even-gap)
superconductivity in this model are singlets: extended $s$-wave
(identity representations of $D_4$ point group symmetry),
$d_{x^2-y^2}$-wave ($B_2$) and $d_{xy}$-wave ($B_1$). We find that,
apart from the standard choice of the pairing state in these channels
as:
$\Delta_{x^2-y^2}=\langle
c_{\uparrow\vec k} c_{\downarrow-\vec k}\rangle \propto \cos k_x - \cos k_y$;
$\Delta_{xy}\propto \sin k_x \sin k_y$,
there is a set of composite operators which satisfy
all the requirements of the symmetry and spin eigenvalues

\begin{eqnarray}
\Delta_{R=\sqrt 2}^{d_{x^2-y^2}}({\bi})&=&
(\vec S_{\bf {i+\hat x}}-\vec S_{\bf {i+\hat y}})\cdot
\vec T_{\bf i, \bf {i+\hat x+\hat y}}  +
(\vec S_{\bf {i-\hat x}}-\vec S_{\bf {i-\hat y}})\cdot
\vec T_{\bf i, \bf {i-\hat x-\hat y}}\\ \nonumber
&+&(\vec S_{\bf {i+\hat x}}-\vec S_{\bf {i-\hat y}})\cdot
\vec T_{\bf i, \bf {i+\hat x-\hat y}}  +
(\vec S_{\bf {i-\hat x}}-\vec S_{\bf {i+\hat y}})\cdot
\vec T_{\bf i, \bf {i-\hat x+\hat y}}\\ \nonumber
\Delta_{R=1}^{d_{xy}}(\bi)&=&
(\vec S_{\bf {i+\hat x+\hat y}}-\vec S_{\bf {i+\hat x-\hat y}})\cdot
\vec T_{\bf i, \bf {i+\hat x}}  -
(\vec S_{\bf {i-\hat x+\hat y}}-\vec S_{\bf {i-\hat x-\hat y}})\cdot
\vec T_{\bf i, \bf {i-\hat x}}\\ \nonumber
&+&(\vec S_{\bf {i+\hat x+\hat y}}-\vec S_{\bf {i-\hat x+\hat y}})\cdot
\vec T_{\bf i, \bf {i+\hat y}}  -
(\vec S_{\bf {i+\hat x}-\hat y}-\vec S_{\bf {i-\hat x-\hat y}})\cdot
\vec T_{\bf i, \bf {i-\hat y}}\\ \nonumber
\Delta_{R=1}^{p_{x}}(\bi)&=&
(\vec S_{\bf {i+\hat x}}-\vec S_{\bf {i-\hat x}})\cdot
(\vec P_{\bf i, \bf {i+\hat y}} + \vec P_{\bf i, \bf {i-\hat y}}),
\\ \nonumber
\label{eq:2}
\end{eqnarray}
\noindent
where $\vec T_{\bf i, \bf j}=\frac{1}{i}c_{\bf i,\sigma} (\sigma^y\vec
\sigma)_{\sigma\sigma^\prime}c_{\bf j,\sigma^\prime}$ and $\vec P_{\bf
i, \bf j}=c_{\bf i,\sigma} (\sigma^x\vec
\sigma)_{\sigma\sigma^\prime}c_{\bf j,\sigma^\prime}$.  We used real
space representation with $\bf \hat x$ and $\bf \hat y$ being unit
vectors in $x$ and $y$ direction respectively. In the last of three
equations in Eq. (2) we present for completeness also $p_x-$
wave triplet $S_z=0$ composite operator. It is interesting to note,
that Cooper pairs in Eqs.(2) lie on symmetry axes of given
symmetries, {\it i.e.} $x^2-y^2=0,~xy=0,~~x=0$, respectively.

The symmetry of the above order parameters is exactly the same as of
the standard operators and corresponds to one of the symmetry
representations on the square lattice.\cite{sym} To describe the
derivation of the composite operators Eq.~(2) for BCS
superconductor we will review the derivation of the composite
operators for odd-gap superconductors \cite{a,bb,ek,c}. The general
form of the two-particle gap function can be written as
\begin{equation}
\Delta_{\bf ij}(t) = \langle T_t c_{{\bf i} \alpha}(t) c_{{\bf j}
 \beta}(0)\rangle\sigma^y_{\alpha\beta}.
\label{eq:3}
\end{equation}
Assuming analyticity of the gap function at small $t\to 0$, the latter
can be expanded for both odd-frequency and BCS channels as \cite{bb}
\begin{eqnarray}
\Delta_{\bf ij}^{\it odd}(t)&=&  \Delta_{\bf ij}^{(1)} t +
{\cal O}(t ^3)\label{eq:3a}\\
\Delta_{\bf ij}^{\it even}(t)&=& \Delta_{\bf ij}^{(0)} +
{1\over 2} \Delta_{\bf ij}^{(2)}t^2 + {\cal O}(t^4).
\label{eq:3b}
\end{eqnarray}
For odd-frequency pairing after taking time derivative
${\partial\Delta_{\bf ij} ^{odd}(t)\over\partial t}\vert_{t=0}=
\Delta^{(1)}_{\bf ij} \propto \langle \left [ H, c_{{\bf i}\alpha}\right ]
c_{{\bf j}\beta}\rangle \sigma_{\alpha\beta}^y$ we arrive at the composite
operator. Here we limit ourselves to the singlet case. The generic
form of the composite operator is always as in Eq.(\ref{eq:1}),
however the details will depend on the Hamiltonian $H$ (for more
details in the case of t-J model see for example \cite{bb}).

To obtain the composite operators for even frequency or BCS pairing
(2), we have to take the second order time derivative
\begin{equation}
\Delta^{(2)}_{\bf ij} =
{\partial ^2\Delta_{\bf ij} ^{even}(t)\over\partial t^2}\vert_{t=0}\propto
\langle \left [H,\left [ H, c_{{\bf i}\alpha}\right ]\right ]
c_{{\bf j}\beta}\rangle \sigma_{\alpha\beta}^y.
\label{eq:4}
\end{equation}
We arrive at the composite operator in the form of Eq.~(2) by taking
for the first commutator the hopping term $H_t = t \sum_{\bf
ij}c_{{\bf i}\sigma}^\dagger c_{{\bf j}\sigma}$ and for the second
commutator the Heisenberg term $H_J=J\sum_{\bf ij} \vec S_{\bf i}\vec
S_{\bf j}$.  The commutator $\left [H_t,c_{{\bf i}\alpha}\right ]$
moves particle to the neighboring site which may lie on the main
symmetry axis. The second commutator with $\left [ H_J, \left
[H_t,c_{{\bf i}\alpha}\right ] \right ]$ produces extra spin operator.
{}From $\left [ \vec S_{\bf i}, c_{{\bf i}\alpha}\right ] = (-{1\over
2})\vec \sigma_{\alpha\nu}c_{{\bf i}\nu}$ we find $\left [ H_J,c_{{\bf
i}\alpha}\right ]= -J\sum_{<{\bf ik}>}\vec S_{\bf k}\vec
\sigma_{\alpha\nu}c_{{\bf i}\nu}$ from where directly follows the
general structure of the operators in Eq.(2) as composite operators of
a Cooper pair with an attached spin operator.  Direct check also
reveals that these composite operators obey the required symmetry
conditions under $D_4$ point group transformations.

In conclusion, we present the list of composite operators for BCS (even
frequency) pairing, using 2D t-J model as an example. The structure of
these composite operators is analogous to the composite operators,
introduced for  the odd-frequency pairing. Important difference
between these operators for odd-frequency and BCS pairing is
that the composite operator for BCS pairing comes from the dressing
of the quasiparticle operator assuming  that standard equal time BCS gap
function is nonzero. Although this dressing might improve overlap
with the ground state, it does not represent new physics.
Situation changes drastically if the usual BCS gap function has very
small or even zero expectation value. In this case the composite BCS
operator  corresponds to the real "pairing" processes. For the
odd-frequency pairing the composite operator indeed represents the
equal time "pairing" in odd-frequency superconductors. As it has been
pointed out in \cite{bb}, the closeness to the instability in the t-J
model helps the composite channel because of a soft spin fluctuations
in the system. Presumably the same holds for the composite BCS
channels in the frustrated correlated systems.

%\vfill

\underline{Acknowledgments} A.B. acknowledges J.R. Oppenheimer
fellowship support.

%\eject


\begin{references}
\bibitem{com} The phase space restrictions will make appearance of
              such a condensate more constraint.
\bibitem{sym} We note that the $d_{x^2-y^2}$ form of the composite
              operator was previosly derived by Poilblanc (ref. \cite{p}).
\end{references}
\end{document}